\documentclass[12pt,a4paper]{article}
\usepackage{graphicx}
\usepackage{times}
\usepackage[pagewise]{lineno}
\title{\bf Ground-based photometric survey to search for the pulsational variability in \texttt{Bp}, \texttt{Ap}, and \texttt{Am} stars}
\author{Daniel Nhlapo$^1$\thanks{e-mail: Daniel.Nhlapo@nwu.ac.za}, Santosh Joshi\thanks{Presenter of the work in 2$^{\rm nd}$ BINA workshop} $^2$, Bruno Letarte$^1$,\\ N. K. Chakradhari$^3$ and S. K. Tiwari$^4$
\vspace{0.5cm}\\
\\
\normalsize $^1$ Centre for Space Research, North-West University, South Africa
\\ 
\normalsize $^2$ Aryabhatta Research Institute of Observational Sciences (ARIES), Manora Peak,\\
Na\normalsize inital 263002, India
\\
\normalsize $^3$ School of Studies in Physics and Astrophysics, Pt. Ravishankar Shukla University,\\
\normalsize 492\,010 Raipur, India
\\
\normalsize $^4$ Siddhi Vinayak Engineering \& Management College, Alwar 301001, Rajasthan, India
}

\date{\mbox{}}
\begin{document}
\maketitle
\setcounter{page}{1001}
\pagestyle{plain}
    \makeatletter
    \renewcommand*{\pagenumbering}[1]{%
       \gdef\thepage{\csname @#1\endcsname\c@page}%
    }
    \makeatother
\pagenumbering{arabic}

%
%
\def\bull{\vrule height .9ex width .8ex depth -.1ex}
\makeatletter
\def\ps@plain{\let\@mkboth\gobbletwobilateral
\def\@oddhead{}\def\@oddfoot{\hfil\scriptsize\bull\quad
``2nd Belgo-Indian Network for Astronomy \& Astrophysics (BINA) workshop'', held in Brussels (Belgium), 9-12 October 2018 \quad\bull}%
\def\@evenhead{}\let\@evenfoot\@oddfoot}
\makeatother
%
%
\def\beginrefer{\section*{References}%
\begin{quotation}\mbox{}\par}
\def\refer#1\par{{\setlength{\parindent}{-\leftmargin}\indent#1\par}}
\def\endrefer{\end{quotation}}
%
%
{\noindent\small{\bf Abstract:} 
We present the analysis of time-series of photoelectric data of a {\tt Bp} star and four new {\tt Ap} stars observed photoelectrically under the Nainital-Cape survey programme. The project was started about two decades ago, aiming to search for new rapidly oscillating {\tt Ap}  stars. The frequency analysis of the time-series of these stars obtained on multiple nights did not reveal any pulsational variability.  In addition to this, we have performed the analysis of time-series differential CCD photometry of the two pulsating {\tt Am} stars HD\,13038 and HD\,13079, where we find some evidence of new periods. To expand and strengthen the ongoing survey work, we propose to build-up a tri-national collaboration of astronomers from India, South Africa and Belgium.\\

{\noindent\small{\bf Keywords:} Survey: Nainital-Cape -- Stars: {\tt Am}, {\tt Ap}, ro{\tt Ap}, $\delta$\,Scuti -- Photometry: high-speed, differential -- variability -- pulsations.

\section{Introduction}

\noindent
Ground-based photometric surveys can be used for the detection and the study of variability in chemically peculiar (CP) stars. One of them is the Nainital-Cape survey (Ashoka et al. 2000), inititated in 1999 to search for new rapidly oscillating {\tt Ap} (ro{\tt Ap}) stars. At that time, only 3 members of this class of pulsating stars were known in the Northern hemisphere. The results obtained from this large ground-based survey are summarized by Martinez et al. (1999, 2001), Girish et al. (2001), Joshi et al. (2003, 2006, 2009, 2010, 2012a, 2012b, 2014, 2016, 2017), and Balona et al. (2013, 2016).

The ro{\tt Ap} stars are a subset of cool {\tt Ap} stars having abnormal abundances of Si and rare earth elements such as Sr, Cr, Eu. They possess a strong global magnetic field of the order of kiloGauss (Hubrig et al. 2012) that supports the formation of patches of various elements around the magnetic poles (Balona et al. 2013). The ro{\tt Ap} stars pulsate in high-overtone\,({\tt n} $\ge$ 20), low-degree ($\ell$ $\le$ 3), non-radial $p$-modes (Shibahashi 1983, Balona et al. 2013) with periods ranging from 6 to 23\,min and peak-to-peak light variations up to 18 millimagnitudes (mmag) in the Johnson $B$-band (Holdsworth et al. 2018).

The other group of A-type CP stars, known as metallic-line {\tt Am} stars, is characterized by an over-abundance of metals (e.g. Fe, Ni, etc.) and an under-abundance of light elements (e.g. Ca, Sc, etc). Some {\tt Am} stars exhibit $\delta$\,Scuti ($\delta$\,Sct) type pulsations with periods ranging from 0.02 to 0.25 days of V-band amplitudes from 0.003 to 0.9 mag (Breger 2000). The $\delta$\,Sct stars are pulsating stars of spectral type between late-A and early-F with a mass range of 1.5--2.5\,M$_{\odot}$ (Chang et al. 2013). They pulsate in low-order radial and/or non-radial $p$- and $g$-modes driven by the opacity mechanism operating in the second helium ionization zone (Houdek et al. 1999, Aerts et al. 2010). The $\delta$\,Scuti stars in binary systems are particularly important as their basic parameters, such as mass and radius, can be determined precisely using the mass-radius relation of binary systems (Soydugan et al. 2016, Alicavus et al. 2017, Mkrtichian et al. 2018). Furthermore, the precise values of these paramaters can be used as input for the modeling of their stellar structure and evolution.  The multi-periodic pulsating stars in such systems are of major importance because multi-mode waves probe different depths of the stellar interior (Joshi $\&$ Joshi 2015). 

In the early 1980s at the South African Astronomical Observatory (SAAO), the Cape survey project was successful in detecting mmag light variations in {\tt Ap}  and {\tt Am} stars (Martinez $\&$ Kurtz 1994). In 1999, almost two decades later, the programme extended to the Northern hemisphere after the Aryabhatta Research Institute of Observational Sciences (ARIES) joined the project. The Nainital-Cape survey's main objective was to find new ro{\tt Ap} stars and pulsating {\tt Am} stars in both the Northern and Southern hemisphere. In this paper we present some of the recent results obtained since the work presented by Joshi et al. (2016).  

\begin{table}[h]
\caption{The basic parameters of the one {\tt Bp}, four {\tt Ap}, and two {\tt Am} stars studied. (Credit: {\tt SIMBAD}). \label{tab-null}}
\scriptsize 
\begin{center}
\begin{tabular}{cccccccccccccccc}
 \hline
 \hline
Star & HD & $\alpha _{2000}$ & $\delta _{2000}$ & m$_{v}$ & SpT &  {\it b-y} & {\it m$_{1}$} & {\it c$_{1}$} & {\it $\beta$} & $\log(T_{\mathrm{eff}})$ & $\log ({L/L_{\odot}})$ & $\Delta t$ &Year\\
 No. & number & ({\tiny hrs\,min\,sec}) & ({\tiny deg\,min\,sec}) & ({\tiny mag}) &  & ({\tiny mag}) & (mag) & ({\tiny mag}) & ({\tiny mag}) & ({\tiny K}) & & ({\tiny hr}) &of Obs. \\
 \hline
 \hline 
1. &  13079  & 02 09 02 & +39 35 32 & 8.76    & Am      & --      & 0.203 & 0.211 & 0.672 & 2.759 & 3.874 & 25.1 & 2016 \\
2. &  13038  & 02 09 30 & +57 57 38 & 8.52 & A4m     & 5.87    & 0.105 & 0.220 & 0.848 & 2.856 & 3.911 & 29.0 & 2016 \\ 
3. &  41089  & 06 00 51 & -42 52 14 & 6.57 & B9IIIp  & --      & --    & --    & --    & 4.179 & 2.49  & 46.1 & 2009 \\
4. &  156869 & 17 22 52 & -52 58 41 & 7.92 & Ap      &  0.041  & 0.165 & 0.990 & --    & 3.980 & 1.77  & 1.8  & 2009 \\
5. &  187473 & 19 51 10 & -27 28 19 & 7.32 & Ap      &  -0.022 & 0.157 & 0.677 & --    & 4.059 & 1.74  & 2.6  & 2009 \\
6. &  189832 & 20 03 35 & -38 51 09 & 6.90 & A6p     &  0.139  & 0.199 & 0.949 & 2.822 & 3.892 & 1.63  & 3.5  & 2009 \\
7. &  212385 & 22 24 38 & -39 07 37 & 6.84 & A2p     &  0.067  & 0.225 & 0.946 & --    & 3.927 & 1.36  & 2.8  & 2009 \\
\hline   
\end{tabular}
\end{center}
\end{table}

\section{Selection of targets and observations}

The effectiveness of any long-term programme relies upon the selection criteria used to prepare a well defined sample of targets to be investigated. Our selection criteria are described by Joshi et al. (2016). Table\,\ref{tab-null} gives the basic physical parameters of the one {\tt Bp}, four {\tt Ap}, and two {\tt Am} candidate stars studied. These parameters were compiled from the {\tt SIMBAD}\footnote{This research has made use of the {\tt SIMBAD} database, operated at CDS, Strasbourg, France (Wenger et al. 2000)} database.  For each star, the columns of Table\,\ref{tab-null} give respectively the HD\,number, right ascension ($\alpha_{2000}$), declination ($\delta_{2000}$), visual magnitude ($m_v$), spectral type (SpT), Str\"omgren photometric indices ($b-y$, $m_1$, $c_1$, $H_{\beta}$), effective temperature ($\log(T_{\mathrm{eff}})$), luminosity ($\log ({L/L_{\odot}})$), duration of the observations ($\Delta$t), and the year when the observations were obtained (Year). We use ``--'' in the case that values are not available in the literature. 

To study short-term periodic variability in stars (P $\leq$ 30\,min), particularly in case of ro{\tt Ap} stars, it is common to use a single channel high-speed photometric technique, which is able to detect short period pulsations with small amplitudes. The basic requirement for this technique is to have high-duty cycle observations, in which the telescope points to the target star without moving to any comparison star to minimize the time loss. To study long-term pulsational variability (P $\geq$ 30 min), as found in $\delta$\,Scuti stars, differential photometry is more suitable. This requires comparisons stars of similar magnitude and color in the same field of view. 

\subsection{Observations of \texttt{Bp} and \texttt{Ap} stars}

For the observations taken from SAAO, we used a high-speed photometric technique. The photoelectric observations of one {\tt Bp} and four {\tt Ap} candidate stars presented here were acquired using a Modular Photometer\,(MP) attached to the 0.5-m telescope at the Sutherland site of SAAO. Each data point comprises of 10-sec observations, which is only a fraction of the expected pulsation periods, through a Johnson $B$-filter. Initially, a single-channel photometric mode with occasional interruptions to measure the sky background was implemented. This exercise of intermediate sky background measurements becomes more relevant when the moon rises and sets, causing a sudden change in the sky counts. A relatively large photometric aperture of 30$^{\prime\prime}$ was chosen to minimize the effects of movement of the star due to seeing fluctuations and tracking errors of the telescope. Each target was monitored continuously for the duration of approximately 1 to 3 hours, depending on the presence of any signature of periodic variation in the raw light curve. During the observing run, if there was evidence of a regular variation in the light curve, as was the case for HD\,41089, then the strategy was to collect data over a maximum duration on the following nights to rule out inherent variation coming from the Earth's atmosphere or from the instrument. In case of HD\,41089, the follow-up observations to confirm any pulsational variability was taken for ten nights. The log of its observations is presented in Table\,\ref{T41089}. 

\subsection{Observations of \texttt{Am} stars}

Out of seven stars discovered as pulsating variables under the Nainital-Cape survey project, only the \texttt{Am} stars HD\,13038 and HD\,13079 were suitable for differential photometry because their field of view of the CCD contains at least two comparison stars of similar brightness. Both of these stars were monitored using a 2K\,$\times$\,2K CCD mounted on the 1.04-m Sampurnanand telescope of ARIES giving a field of view of 13 arcmin\,$\times$\,13 arcmin. The left and right panels of Fig.\,\ref{comparison} show identification charts of HD\,13038 and 
HD\,13079 on a same scale, respectively, where the location of the target and comparison stars are marked. Table\,\ref{tab-2stars-log} gives a log of the photometric observations obtained for these stars.

\begin{figure}[ht]
\centering
\includegraphics[width=0.45\textwidth]{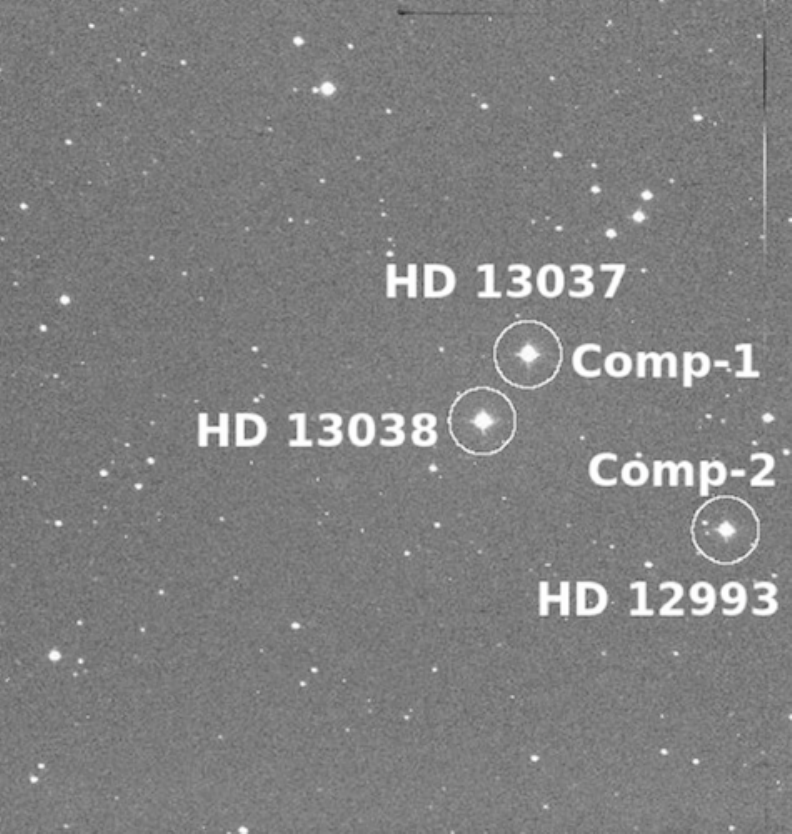}
\includegraphics[width=0.45\textwidth]{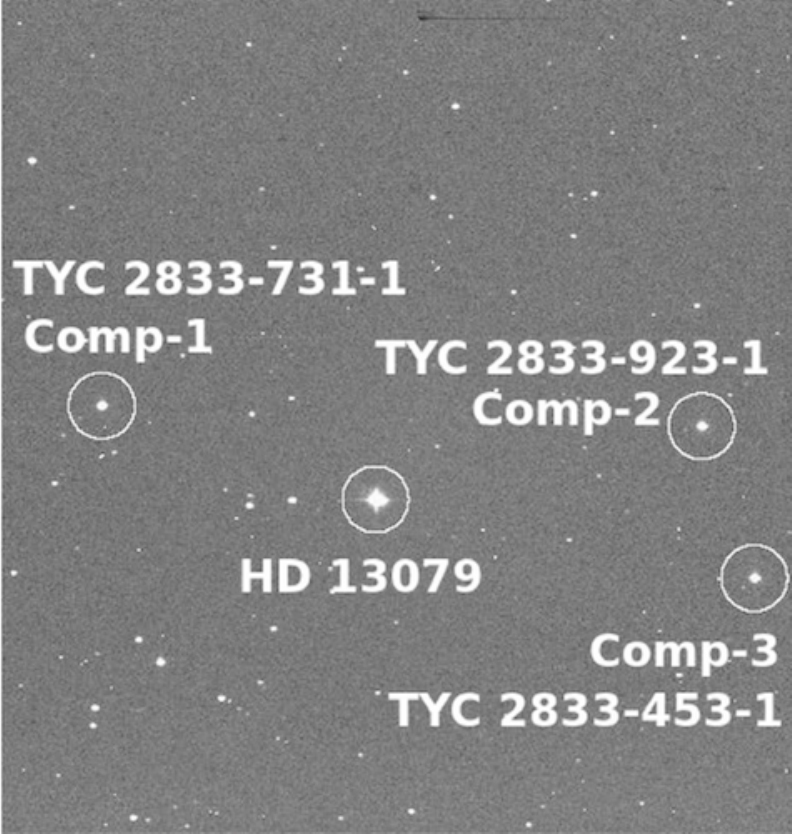}
\caption{The identification charts of HD\,13038 (left) and HD\,13079 (right) with their comparison stars in a field of view of 13\,arcmin\,$\times$\,13\,arcmin.} 
\label{comparison}
\end{figure}

\begin{table}[!h]
\caption{Summary of the photometric observations of HD\,13079 and HD\,13038 obtained with the 1.04-m Sampurnanand telescope of ARIES.\label{tab-2stars-log}}
\small
\begin{center} 
\begin{tabular}{|cc|cc|}
\hline
\multicolumn{2}{|c}{\bf HD\,13079}&\multicolumn{2}{|c|}{\bf HD\,13038}\\ 
HJD & $\Delta t$& HJD & $\Delta t$\\
    & (hrs)     &     & (hrs)     \\ 
\hline
2450765 &  2.0  & 2451146 &  3.8  \\
2450768 &  2.5  & 2451147 &  4.6  \\
2451197 &  3.1  & 2451151 &  3.2  \\
2451198 &  2.2  & 2451152 &  6.7  \\
2451199 &  2.6  & 2452282 &  3.6  \\
2454013 &  4.5  & 2452287 &  1.6  \\
2454016 &  8.2  & 2452295 &  1.2  \\
        &       & 2452611 &  1.1  \\  
        &       & 2452934 &  1.1  \\
        &       & 2457387 &  2.1  \\ 
\hline 
Total   & 25.1  & Total   & 29.0  \\ 
\hline 
\end{tabular}
\end{center} 
\end{table}

\section{Data reductions and analysis}

The data reduction process of the fast photometry starts with a visual inspection of the raw light curves to identify and remove outliers, followed by a correction for coincidence counting losses, a subtraction of the interpolated sky background, and a rectification of the mean atmospheric extinction. The time of midpoint of each observation was converted into the Heliocentric Julian Date (HJD) with an accuracy of 10$^{-5}$ day ($\sim$1 second).

The CCD images from ARIES obtained for the two pulsating {\tt Am} stars HD\,13079 and HD\,13038 were processed using standard tasks available in {\sc iraf}\footnote {{\sc iraf} is distributed  by the National Optical Astronomy Observatories, which are operated by the Association of Universities for Research in Astronomy, Inc., under cooperative agreement with the National Science Foundation.}. The images were corrected using a master bias frame prepared as the median of the good-quality bias frames taken throughout the night. Normalized flat-field images of the twilight sky were used to correct for the pixel-to-pixel sensitivity. Finally, differential CCD photometry was performed to measure the magnitude of the target relative to the comparison stars to eliminate sky-transparency variation.

The time-series data of each observing run from both SAAO and ARIES were subjected to frequency analysis. A fast algorithm based on the Lomb-Scargle (LS) periodogram (Lomb 1976, Scargle 1982) for unequally spaced data was used to extract any periodic variation present in the time-series data. To search and extract peaks due to pulsation, the frequency with the highest amplitude was removed by subtracting the corresponding sinusoid from the data. This process was repeated until the noise level in the periodogram of the residuals was reached.


\section{Results}

The left and right panels of Fig.\,\ref{fig_1} show time-series and corresponding periodograms of the four {\tt Ap} stars listed in Table\,\ref{tab-null}. Upon examining these figures, it is noticeable that the light variation is probably due to the sky-transparency. The left and right panels of Fig.\,\ref{lc-hd41089} show the time-series and corresponding periodograms of the {\tt Bp} star HD\,41089 obtained on 11 different nights. The frequency solution of time-series collected on different epochs is listed in columns\,3 and 4 of Table\,\ref{T41089}. It is evident from this table that the dominant low frequencies could be due to long period sky-transparency variations.
Therefore, differential photometry is suggested to rule out any contribution from sky-transparency variations. Based on the available observations, we could not detect ro{\tt Ap} like pulsational variability in any of the 5 candidate stars (HD\,187453, HD\,189832, HD\,212385, HD\,156869, HD\,41089).

\begin{figure}
\centering
\includegraphics[width=0.9\textwidth]{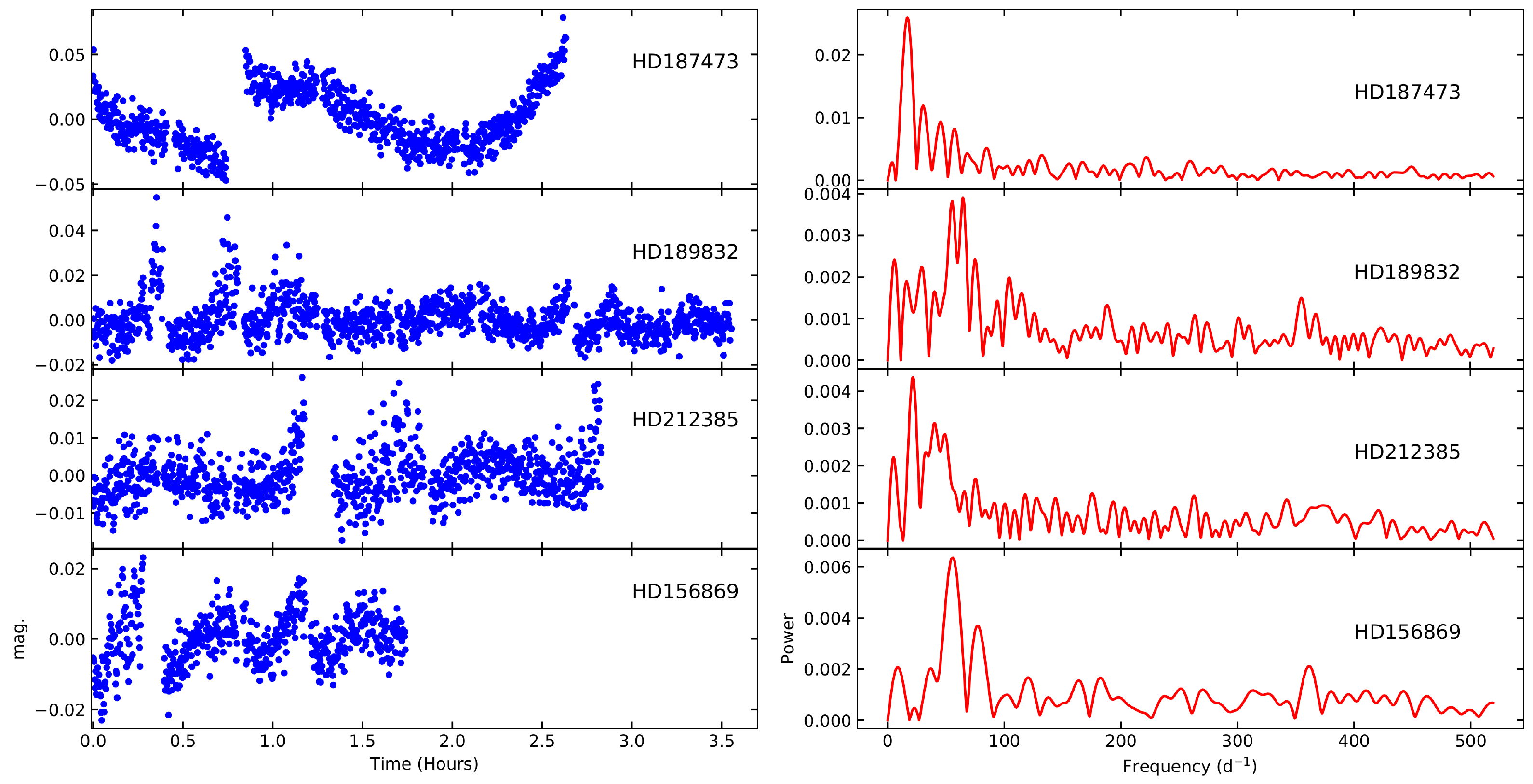}
\caption{The photo-electric light curves and corresponding periodograms of four candidate {\tt Ap} stars observed from SAAO.} 
\label{fig_1}
\end{figure}

\begin{figure}
\centering
\includegraphics[width=0.9\textwidth]{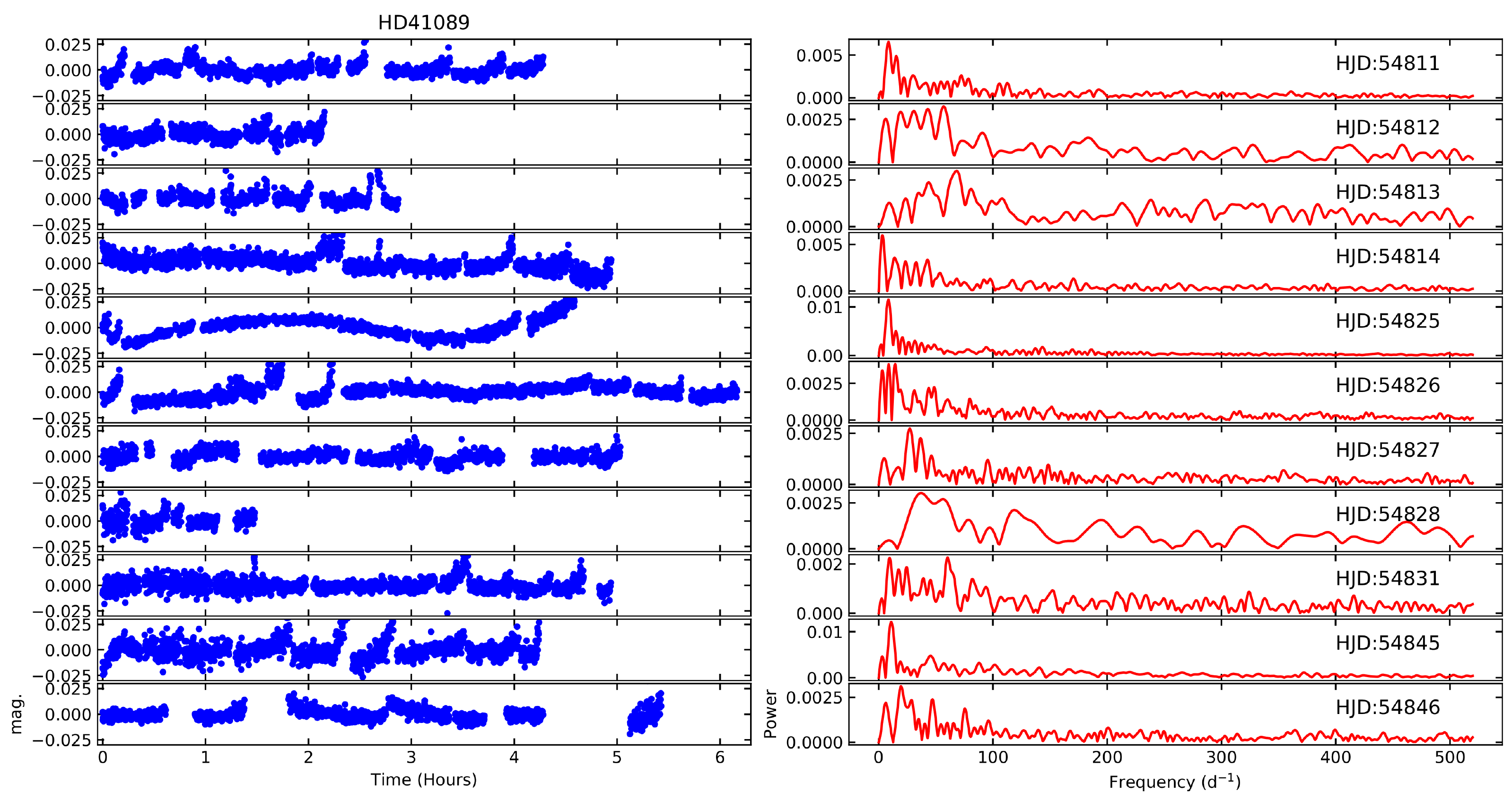}
\caption{The photo-electric light curves and corresponding periodograms of HD\,41089 obtained on different nights.} 
\label{lc-hd41089}
\end{figure}

\begin{table}[!h]
\centering
\parbox{.40\linewidth}{
\caption{The frequency solution of HD\,41089.\label{T41089}}
\small
\begin{center}
\begin{tabular}{|cccc|}
\hline
HJD     & $\Delta$t & Frequency & Power \\ 
        & (hrs)     & (d$^{-1}$) &       \\
\hline
2454811 &       4.3 &     0.100 & 0.635 \\ 
2454812 &       2.2 &     6.781 & 0.529 \\ 
2454813 &       2.9 &    15.141 & 0.098 \\ 
2454814 &       4.9 &     0.100 & 0.359 \\ 
2454825 &       4.6 &     0.100 & 0.953 \\ 
2454826 &       6.2 &     0.100 & 0.185 \\
2454827 &       5.0 &    10.596 & 0.291 \\
2454828 &       1.5 &     0.100 & 0.394 \\ 
2454831 &       4.9 &     4.960 & 0.430 \\ 
2454845 &       4.2 &     4.627 & 0.835 \\ 
2454846 &       5.4 &    13.365 & 0.158 \\ 
\hline
Total   &      46.1 &   0.187   & 0.0283  \\
\hline 
\end{tabular}
\end{center}
}
\parbox{.05\linewidth}{~~~}
\parbox{.50\linewidth}{
\caption{Known and new frequencies of HD\,13079 (top) and HD\,13038 (bottom). The signal-to-noise ratio (SNR) listed in column\,2 is the ratio of the power of the dominant frequency to the power averaged over all peaks in the frequency spectrum. \label{tab-2stars1}}
\small
\begin{center} 
\begin{tabular}{|lcl|}
\hline
\multicolumn{3}{|c|}{\bf HD\,13079}                     \\
Frequency             &    SNR & Reference              \\
 (d$^{-1}$)            &        &                        \\
\hline 
 18.383\,$\pm$\,0.006 & 20.823 & new                    \\
~~9.737\,$\pm$\,0.006 & 10.480 & new                    \\
 18.799\,$\pm$\,0.006 &~~8.433 & new                    \\ 
 19.759               &    -   & Martinez et al. (2001) \\
 19.4090              &    -   & Smalley et al. (2011)  \\
 19.4552              &    -   & Smalley et al. (2011)  \\
 24.7550              &    -   & Smalley et al. (2011)  \\
\hline
\hline      
\multicolumn{3}{|c|}{\bf HD\,13038}                     \\
Frequency             & SNR    & Reference              \\
 (d$^{-1}$)            &        &                        \\
\hline 
 49.147\,$\pm$\,0.003 & 27.414 & new                    \\
 52.942\,$\pm$\,0.003 & 21.354 & new                    \\
 50.196               &    -   & Martinez et al. (2001) \\
 39.811               &    -   & Martinez et al. (2001) \\
\hline   
\end{tabular}
\end{center}
}
\end{table}

\begin{figure}[!h]
\centering
\includegraphics[width=0.999\textwidth]{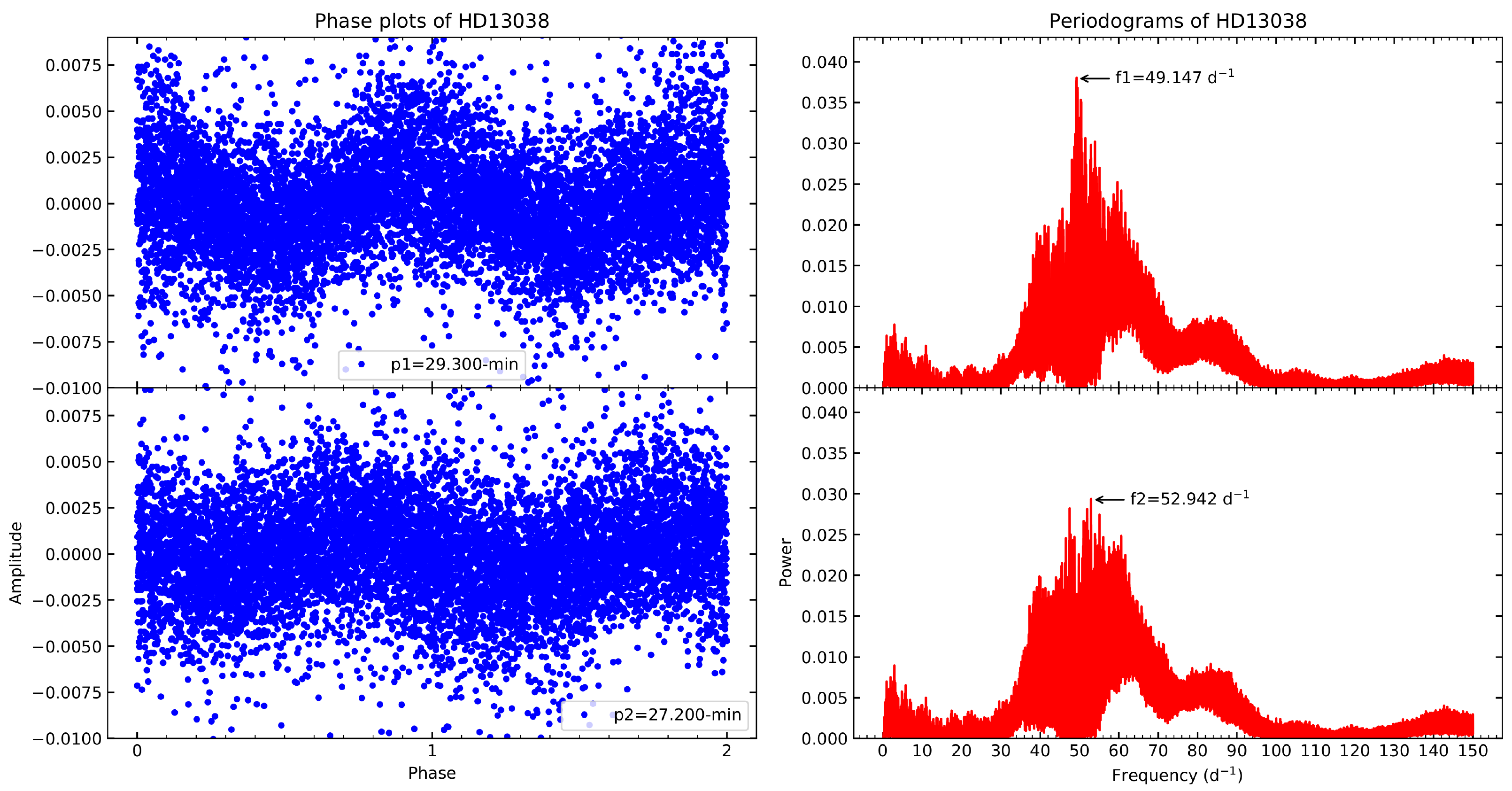}
\caption{Differential phase-light curves and corresponding periodograms of HD\,13038.
\label{lc13038}}
\end{figure}

\begin{figure}[!h]
\centering
\includegraphics[width=0.999\textwidth]{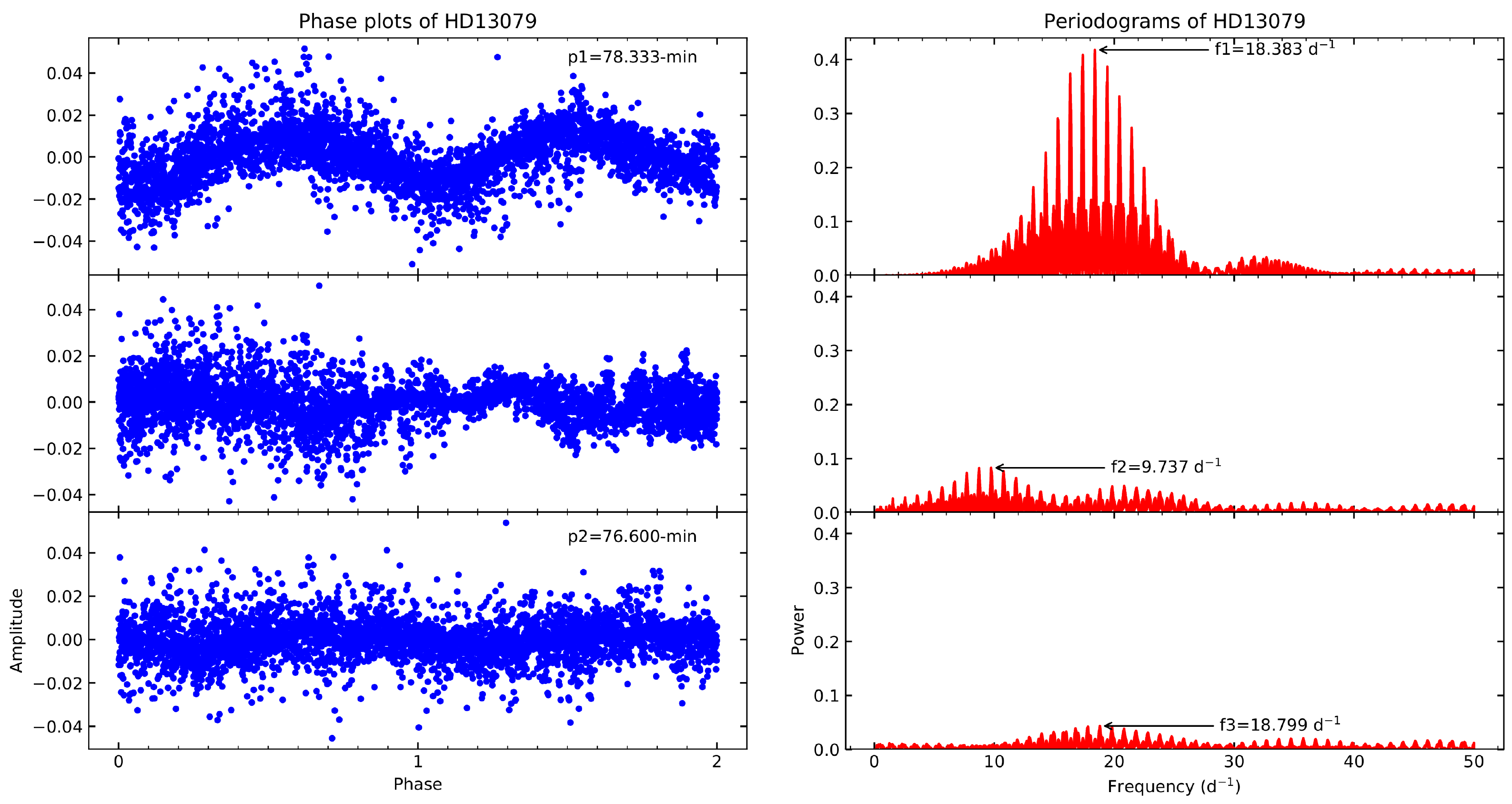}
\caption{Differential phase-light curves and corresponding periodograms of HD\,13079.
\label{lc13079}}
\end{figure}

The left panels of Fig.\,\ref{lc13038} display the phased time-series of the combined data of HD\,13038 obtained through the Johnson-$B$ band while the periodograms are shown in a right panels of this figure. The periodogram of the original data shows a dominant frequency at f$_{1}$\,=\,49.147\,d$^{-1}$. This frequency has never been reported before for this star. After prewhitening the dominant frequency, the Fourier analysis of the residual time-series reveals f$_{2}$\,=\,52.942\,d$^{-1}$ as the second frequency. Similarly, the left panels of Fig.\,\ref{lc13079} show the phased time-series of the combined data of HD\,13079 obtained through the Johnson $B$-band. The periodograms of the 
time-series without and with pre-whitening of the subsequent frequencies (from top to bottom) are shown on the right panels of this figure. The periodogram of the original data shows a dominant frequency f$_{1}$\,=\,18.383\,d$^{-1}$, which was not reported before. Prewhitening of this dominant frequency reveals two additional frequencies: f$_{2}$\,=\,9.737\,d$^{-1}$ and f$_{3}$\,=\,18.799\,d$^{-1}$. The errors of the new frequencies in Table\,\ref{tab-2stars1} were calculated with Monte Carlo simulations in the software package \small{\tt PERIOD04} (Lenz \& Breger 2005).

\section{Discussion and conclusion}

We have presented time-series and periodograms for one {\tt Bp} star (HD\,41089) and four {\tt Ap} stars (HD\,156869, HD\,187473, HD\,189832, HD\,212385) to search for pulsational variability. Based on data acquired for each star, we could not draw any firm conclusion on their variability. Since the pulsating {\tt Ap} and {\tt Am} stars exhibit amplitude modulation due to either rotation or interference from close pulsational frequencies, the non-variability could be explained by a coincidence in the timing of observations. Therefore, HD\,41089 was monitored more extensively on 11 different nights. Based on these observations, we could not ascertain any conclusive variability. 

We have also performed differential CCD photometry to investigate the nature of the two known pulsating {\tt Am} stars HD \,13038 and HD\,13079 discovered under the Nainital-Cape survey. Our analysis shows that HD\,13079 exhibits three new frequencies: f$_{1}$\,=\,18.383\,d$^{-1}$, f$_{2}$\,=\,9.737\,d$^{-1}$, and f$_{3}$\,=\,18.799\,d$^{-1}$. Note that the frequency 19.4090\,d$^{-1}$, as reported by Smalley et al. (2011), is close to the first harmonic of f$_{2}$. Similarly, it appears that two new pulsational frequencies are present in the time-series data of HD\,13038: f$_{1}$\,=\,49.147\,d$^{-1}$ and f$_{2}$\,=\,52.942\,d$^{-1}$.  
Finding additional frequencies in variable stars potentially enables us to probe additional regions of their interiors. The non-detection of previously observed frequencies of these two stars could mean that they are below noise level of the data analysed. Another possible explanation is intrinsic amplitude modulation as observed for $\delta$\,Scuti stars (Bowman et al. 2016). For the comparison, we also added the frequencies reported by Martinez et al. (2001) and Smalley et al. (2011) in Table\,\ref{tab-2stars1}. On comparing them with those that we found and by taking into account the errors that we have calculated, it is possible that some of our frequencies are 1\,d$^{-1}$ alias frequencies of those reported before.

The Belgo-Indian Network for Astronomy and astrophysics (BINA) is a collaborative project involving dozens of astronomers from India and Belgium working in different areas of astrophysics. The network is gradually expanding to other nations. Our attempts to make a multi-nation collaboration have been recently rewarded in terms of a joint publication by astronomers from India, Belgium and South Africa (Chowdhury et al. 2018), laying the foundations for a tri-national collaboration. With the development of new observational facilities at Devasthal (Stalin et al. 2001, Kumar et al. 2018) operated by ARIES (Nainital), it is expected that the data obtained using new technology telescopes in India equipped with state-of-art instruments can play a crucial role in future achievements in asteroseismology of CP stars.

\section*{Acknowledgements}

The work presented here is carried out under the international projects: the Indo-South African DST/INT/SA/P-02 \& INT/SAFR/P-3(3)/2009 and Indo-Belgium ‘BINA’ projects DST/INT/Belg/P-02 \& BL/11/IN07. We thank Daniel Holdsworth for a fruitful discussion on interpretaion of our results and continuous assistance on error analysis.

%
%

\footnotesize
\beginrefer

\refer Aerts C., Christensen-Dalsgaard J., Kurtz D.~W.  2010, Astroseismology, Springer-Verlag, Berlin
   
 \refer Alicavus F.~K., Soydugan F. 2017, AIP Conf. Proc., 1815, 080004
  
 \refer Ashoka B.~N., Seetha S., Raj E. et al. 2000, BASI, 28, 251

 \refer Balona L.~A., Catanzaro G., Crause L. et al. 2013, MNRAS, 432, 2808

 \refer Balona L.~A., Engelbrecht C.~A., Joshi Y.~C. et al. 2016, MNRAS, 460, 1318
 
 \refer Bowman D., Kurtz D.~W., Breger M. et al. 2016, MNRAS, 460, 1970B

 \refer Breger M.  2000, ASP Conf. Ser., 210, 3
 
 \refer Chang S.~W., Protopapas P., Kim D.~W. et al. 2013, ApJ, 145, 132 
 
 \refer Chowdhury S., Joshi S., Engelbrecht C.~A. et al. 2018, Ap$\&$SS, 363, 260
 
 \refer Girish V., Seetha S., Martinez P. et al. 2001, A$\&$A, 380, 142
   
 \refer Holdsworth D.~L., Kurtz D. W., Saio H. et al. 2018, MNRAS, 473, 91 
  
 \refer Houdek G., Balmforth N.~J., Christensen-Dalsgaard J. et al. 1999, A$\&$A, 351, 582 
  
 \refer Hubrig S., Kurtz D.~W., Sch\"oller M. et al. 2012, ASP Conf. Ser., 462, 318 
 
 \refer Joshi S., Girish V., Sagar R. et al. 2003, MNRAS, 344, 431 
 
 \refer Joshi S., Mary D.~L., Martinez P. et al. 2006, A$\&$A, 455, 303
 
 \refer Joshi S., Mary D.~L., Chakradhari N.~K. et al. 2009, A$\&$A, 507, 1763 
 
 \refer Joshi S., Ryabchikova T., Kochukhov O. et al. 2010, MNRAS, 401, 1299 
  
 \refer Joshi S., Semenko E., Martinez P. et al. 2012a, MNRAS, 424, 2002
  
 \refer Joshi S., Joshi Y. C. 2015, A$\&$A, 36, 33
 
 \refer Joshi S., Martinez P., Chowdhury S. et al. 2016, A$\&$A, 590, A116

 \refer Joshi S., Semenko E., Moiseeva A. et al. 2017, MNRAS, 467, 633 
 
 \refer Joshi Y.~C., Joshi S., Kumar B. et al. 2012b, MNRAS, 419, 2379

 \refer Joshi Y.~C., Balona L.~A., Joshi S. et al. 2014, MNRAS, 437, 804
 
 \refer Kumar B., Omar A., Maheswar G. et al. 2018, BSRSL, 87, 29
 
 \refer Lenz P., Breger M. 2005, CoAst, 146, 53

 \refer Lomb N.~R. 1976, Ap$\&$SS, 39, 447 
  
 \refer Martinez P., Kurtz D.~W. 1994, MNRAS, 271, 129
  
 \refer Martinez P., Kurtz, D.~W., Ashoka B.N. et al. 1999, MNRAS, 309, 871
 
 \refer Martinez P., Kurtz D.~W., Ashoka B.N. et al. 2001, A$\&$A, 371, 1048 
 
 \refer Mkrtichian D.~E., Lehmann H, Rodriguez E. et al. 2018, MNRAS, 475, 4746
  
 \refer Scargle J.~D. 1982, ApJ, 263, 835

 \refer Shibahashi H. 1983, ApJL, 275, L5.

 \refer Smalley B., Kurtz D.~W., Smith A.~M.~S. et al. 2011, A$\&$A, 535, A3
 
 \refer Soydugan E., Soydugan F., Alicavus F.~K. et al. 2016, NewA, 46, 40
 
 \refer Stalin C.~S., Sagar R., Pant P. et al. 2001, BASI, 29, 39

 \refer Wenger M., Oschsenbein F., Egret D. et al., 2000, A$\&$AS, 143, 9

\endrefer

\end{document}